**Title:**

# Deep Learning for Blood-Brain Barrier Permeability Prediction: From Discriminative Models to Mechanism-Aware Design

**Author:**

Zihan Yang[1], Yuchen Xiao[2]*

**Affiliations:**

[1]Department of Biosciences, Durham University, Durham DH1 3LE, United Kingdom

[2]School of Life Sciences, Westlake University, Hangzhou, China

*Correspondence: xiaoyuchen@westlake.edu.cn

**Abstract:**

Predicting whether a molecule can cross the blood–brain barrier (*BBB*) is a key step in early-stage neuro-pharmaceutical design, directly influencing the efficiency and success rate of drug development. Traditional methods based on physicochemical properties are prone to systematic misjudgements due to their reliance on previous empirical evidence. Early machine learning (ML) models, although data-driven, often suffer from limited capacity, poor generalization, and insufficient interpretability. In recent years, more advanced models have become essential tools for predicting BBB permeability and guiding related drug design, owing to their ability to simulate molecular structures and capture complex biological mechanisms. This article systematically reviews the evolution of this field—from deep neural networks to graph-based structural modelling—highlighting the advantages of multi-task and multimodal learning strategies in identifying mechanism-related features. We further explore the emerging potential of generative models and causal inference methods for integrating permeability prediction with mechanism-aware drug design. Nowadays, ML-based BBB crossing prediction is in the critical transition from mere discriminative classification toward

structure–function modelling from a mechanistic perspective. This paradigm shift provides a methodological progression and future roadmap for the integration of AI into neuropharmacological development.

**Key points**

• Blood–brain barrier (BBB) crossing prediction has traditionally been framed as a binary classification task, ignoring the underlying transport mechanisms.

• Early machine learning and deep neural network models improve predictive performance but remain circumscribed within correlation-driven representations.

• Graph neural networks enhance molecular structural expressiveness and local interpretability, providing a foundation for mechanism-aware modeling.

• Integrating multimodal learning, generative models, and causal reasoning enables the transition from binary prediction to mechanism-guided molecular design.

• Future BBB modeling frameworks should be mechanistically faithful, intervention-sensitive, and capable of producing actionable insights.

Main Text:

**Overview of Blood-Brain Barrier Permeability**

The development of central nervous system (*CNS*) drugs has long been suffered from high costs and exceptionally low clinical success rates[1]. Between 2009 and 2018, the clinical failure rate reached 84.6%[2]. One of the principal barriers lies in the blood–brain barrier (BBB)—a highly selective and tightly regulated physiological interface that prevents over 98% of small molecules and nearly all biologics from entering the brain, resulting in frequent drug targeting failure of otherwise promising candidates[3–5]. Beyond its physical restrictiveness, the BBB involves a network of complex transport mechanisms that are still insufficiently understood, including passive diffusion, carrier-mediated uptake, and energy-dependent efflux, further hurdling CNS delivery design[6]. Accordingly, accurate early-stage prediction of BBB permeability is crucial for weeding out non-viable compounds during screening, effectively reducing synthesis and evaluation costs[7]. More importantly, integrating specific mechanistic and pathway semantics could enhance the current state of the art such that the models may yield critical and generalizable insights for identifying the "Achilles' heel" of the BBB.

Although deep learning and related AI approaches have been introduced to BBB modelling tasks in recent years[8], the majority of existing models continue to formalize the task as a standard classification or regression problem. This workflow—driven by deployment convenience and the industrial demands for high-throughput compatibility[9]—typically involves learning from molecular descriptors or graph structures to predict whether a compound can "cross" or not. While this method yields reasonably strong predictive performance (reported ROC–AUC values typically ranging from ~0.75 to 0.90 across common datasets) and is well-suited to screening pipelines, it reduces complex, multi-stage physiological processes to one atomic label, obscuring the underlying mechanistic heterogeneity. The resulting models thus often lack interpretability, failing to align semantically with the biological processes of crossing the BBB. Consequently, their outputs provide limited explanatory power, barely aiding structure-based design and optimization. These factors together lead to the bottlenecks of pure discriminative models. To improve the landscape, recent studies have begun to explore mechanism-aware modelling, path-variable integration, and generative approaches for permeability-oriented molecular design[10]—though these efforts remain in their early stages. What accompany the paradigm shift are changes in the models' underlying architectures: from global feature-driven deep neural networks (*DNNs*) to more structurally expressive graph neural networks (*GNNs*)[11,12] and mechanism-grounded generative frameworks. This review examines the evolution of deep learning models for BBB permeability prediction through the lens of task definition transformation. We assess models based on their capacity for structural representation, mechanistic alignment, and actionable feedback, and explore how future frameworks might improve semantic fidelity and practical utility.

**Box 1. Mechanisms governing drug permeability across the BBB**

The blood–brain barrier is formed by tightly connected endothelial cells, supported by pericytes and astrocytes that coordinate barrier integrity and signaling[13]. BBB permeability therefore emerges from multicellular interactions rather than a single membrane property.

Major transmembrane pathways include:
(i) passive diffusion driven by physicochemical properties (molecular weight, lipophilicity, polar surface area);
(ii) carrier-mediated transport via nutrient transporters (e.g., GLUT1, LAT1)[7], which define subtype-specific uptake constraints;
(iii) receptor-mediated transport for macromolecules (e.g., transferrin/insulin receptors);
(iv) adsorptive uptake by surface charge;
(v) active efflux dominated mediated by ABC transporters (e.g., P-gp, BCRP)[14].

BBB permeability is dynamically regulated by biological (development, aging, sex hormones) and pathological (neuroinflammation, neurodegeneration) states. Moreover, experimental context (e.g. in vitro vs. in vivo, co-administered inhibitors) also influences transporter expression and pathway balance[15]. A same molecule may display distinct permeability under different conditions, complicating cross-system generalization.

Addressing the complexity requires a mechanism-aware modelling approach that: (i) decomposes permeability into its constituent biological processes; (ii) links transport features to experimentally actionable labels; and (iii) conditions predictions on physiological context—ultimately generating outputs that can directly guide CNS drug design.

## Discriminative modelling based on static vector inputs: performance gains amid mechanistic blindness

### *I. The binary classification framing: historical roots of BBB permeability prediction*

The prediction of BBB permeability was historically framed as a binary classification task: determining whether a compound could cross the BBB or not[5]. This approach arose from multiple converging factors shaped by early-stage drug discovery and molecular modelling[13].

First, experimental limitations played a central role. Between the late 20th and early 21st centuries, available methods—such as animal models, *in vitro* permeability assays, and logBB partitioning measurements—primarily yielded coarse, binary outcomes[7]. For instance, classic logBB assays only provided rough estimates of brain-to-blood concentration ratios, without weighing the contributions of passive diffusion versus active transport[9], or detailing molecular interactions with transporters[16]. While modern techniques like single-molecule imaging, transporter-specific knockout models, and microfluidic BBB-on-chip systems offer richer mechanistic insights[17], such fine-grained data were inaccessible when the original task frameworks were established.

Second, the state of the art of computational tools influenced task definition. Early rule-based systems, physicochemical heuristics, and classical machine learning (*ML*) methods were inherently aligned with static, fixed-feature classification problems[18]. No mainstream frameworks were capable of processing graph-structured or sequential molecular data at that time; models instead relied on precomputed descriptors such as molecular weight, log*P* (i.e., logarithm of the octanol–water partition coefficient), and PSA (i.e., polar surface area) to optimize decision boundaries[19]. These approaches were effective for binary discrimination but lacked the capacity to generate multilevel or dynamic outputs[20].

Last and most importantly, practical industrial needs decisively shaped the task's formulation. Pharmaceutical companies typically need to quickly eliminate non-viable candidates from thousands to tens of thousands of compounds to be screened, in order to reduce screening costs and accelerate lead advancement. As a consequence, rapid binary classification outputting succinct yes-or-no results best served this goal. While mechanistic insights were scientifically valuable, they were deprioritized in favor of throughput and efficiency in early-stage filtering pipelines.

The binary framing offered distinct advantages: it lowered data demands[20] and provided a simple framework for benchmarking algorithmic performance[21]. However, while it accelerated certain early-stage pharmaceutical processes, its feature engineering including significant structural simplifications would later hinder mechanistic modelling and design-oriented applications—limitations that are addressed in subsequent sections.

## *II. Rule-Based and classical ML approaches: early gains and structural constraints*

In the development of BBB permeability prediction tasks, rule-based systems and classical ML approaches were the technological backbone in the early days[21]. Rule-based methods primarily relied on empirically derived physicochemical thresholds—such as Lipinski's rule of five, Veber's oral bioavailability criteria, molecular weight cutoffs, log*P*, and PSA—to link BBB permeability potential with chemical heuristics. These approaches provided simple and interpretable filters widely used in early-stage drug screening to rapidly eliminate undesirable compounds, especially in settings where only limited experimental data were available[17,22–24].

With the expansion of computational power and data availability, classical *ML* methods such as support vector machine (*SVM*), random forest (*RF*), k-nearest neighbours (*KNN*), and decision tree began to show up in BBB prediction[25,26]. These models typically operated on fixed-input features, including circular molecular fingerprints like the extended-connectivity fingerprints (ECFP) and physicochemical descriptors[20], and used statistical learning to optimize decision boundaries[27]. Compared to rule-based systems, they could capture more complex nonlinear relationships and significantly improved predictive accuracy; for example, early SVM-based models achieved notable gains in the area under the curve (*AUC*) of cross-validation and other efficacy metrics on BBB benchmark datasets. Despite the data scale and computational constraints of the time, classical *ML* remarkably advanced and accelerated early BBB prediction pipelines.

However, these approaches soon revealed inherent structural limitations. First, they were heavily dependent on manually engineered input features, which—while effective for simple chemical space exploration—failed to reflect the hierarchical nature of molecular structures or captured only local topological patterns[28]. Second, they lacked end-to-end learning capabilities, rendering it impossible to automatically extract richer patterns (e.g., graph connectivity, three-dimensional conformations, dynamic transport behaviours, etc.) from raw molecular data[19]. Third, they focused solely on input–output mappings without engaging with the underlying biochemical mechanisms that govern membrane permeation, transporter interactions, and/or energy barriers. In short, early ML techniques maximized classification performance but offered little mechanistic insight or structural interpretability[29].

Importantly, these limitations were not merely technical constraints but were tightly coupled to the prevailing task definition. If the problem was framed as a binary classification task relying on static molecular fingerprints and physicochemical descriptors, classical ML models remained the most practical and efficient solution. Yet, as data complexity increased and research priorities evolved, it became clear that fixed-input, shallow-pattern classification frameworks could no longer meet the growing demand for more complicated and comprehensive modelling objectives[30]—especially those related to mechanistic representation, structural characterization, and multilevel predictions[31].

Against this backdrop, deep learning (DL) methods entered the BBB prediction landscape[32]. Unlike classical ML, deep neural networks (DNNs) offered multilayered feature extraction, hierarchical abstraction, and powerful capacity for learning complex data patterns directly from raw or minimally processed inputs[33]. Their arrival marked not just a performance upgrade but a conceptual shift: from shallow, feature-engineered classification toward deep, end-to-end structural modelling—laying the groundwork for subsequent methodological advances in the field.

## *III. Deep neural networks: feature extraction power and mechanistic limitations*

In BBB permeability prediction tasks, DNNs are currently among the best-performing and most widely applied deep learning tools[18]. Compared to traditional ML methods such as SVM and RF, DNNs offer powerful nonlinear modelling capabilities and multilayer feature extraction, allowing the automatic learning of complex input–output relationships from large-scale molecular datasets[34]. This makes DNNs the standard baseline models in the BBB permeability prediction field[35] and one of the core methods adopted in both academic research and industrial applications[9].

The distinctive strength of DNNs lies in two aspects: the input data and the network architecture. On the input side, DNNs commonly integrate hybrid molecular representations, such as ECFP, *MACCS* key fingerprints, physicochemical descriptors (e.g., log*P*, molecular weight, hydrogen bond donor/acceptor counts, etc.), and SMILES-based sequence encodings[19]. These representations condense chemical information at multiple levels, capturing not only global molecular properties but also local features such as fragment combinations and neighbourhood environments[36]. For example, ECFP generates multilayer fragment features based on atomic neighbourhoods, while SMILES encodes molecular structures as character sequences, making them suitable for sequence-based modelling[37]. On the architecture side, DNNs primarily rely on multilayer perceptrons (*MLPs*), leveraging techniques such as *ReLU* activation, dropout and batch normalization to efficiently model complex nonlinear relationships in high-dimensional feature spaces. In specific tasks, especially those involving SMILES inputs, some studies have explored using convolutional neural networks (CNNs) to extract local sequence patterns[38] or recurrent neural networks (RNNs) to capture long-range dependencies[39], though these approaches are less common compared to MLPs. Overall, the strength of DNNs comes from the synergy between information-rich input features and highly adaptable network architectures, allowing them to process diverse molecular data robustly and effectively.

DNNs have demonstrated outstanding performance across multiple concrete studies. For example, Jing et al. applied ECFP fingerprints with an MLP architecture to ADME-related datasets for BBB permeability prediction, achieving an AUC of ~0.90—significantly outperforming RF- and SVM-based methods[40]. Another study using the *BBBP* dataset employed SMILES encodings combined with a hybrid CNN–DNN multitask model, showing an approximate 6% improvement in accuracy over single-task DNNs, along with enhanced generalizability. Additionally, in industrial high-throughput screening scenarios, there are successful cases where large-scale DNN-based prediction platforms have been used to rapidly screen millions of compounds for potential BBB permeability, significantly reducing the time and cost of experimental validation. These real-world examples highlight not only the academic strength of DNN models but also their practical applicability and scalability in drug development pipelines.

However, DNNs also present important limitations, particularly regarding mechanistic interpretation and task extensibility[41]. Although DNN models substantially improve predictive performance compared with earlier ML approaches, they remain primarily correlation-driven, optimizing for label fitting rather than establishing explicit mechanistic variables or causal pathways[42,43]. Post hoc interpretation tools can only provide approximate estimates of feature importance and do not explain why such features influence BBB transport or how they mechanistically contribute to permeability[29,42]. Moreover, DNN architectures remain constrained by their reliance on predefined molecular encodings, limiting their ability to incorporate richer structural and dynamic information such as molecular topology, three-dimensional conformations, and transport processes across the BBB[44]. As a result, DNN models are prone to shortcut learning and may generalize poorly to novel molecular scaffolds or functional perturbations[45].Overall, DNNs remain powerful predictive tools but offer limited insight into the structural and mechanistic determinants of BBB permeability.

**Graph-based structural modeling as a precursor to mechanistic awareness: From semantic mismatch to expressive enhancement**

***I. Enhancing model interpretability: from attribution analyses to structural embedding***

In BBB permeability prediction tasks, model interpretability is not an optional add-on, but a core requirement directly tied to drug development. It is not enough to simply predict whether a molecule can cross the BBB; drug developers also need to understand how it happens, which molecular features or mechanisms are responsible, and how structural modifications can improve permeability[46]. However, the DNNs commonly used today are fundamentally black-box models: they provide statistical correlations between input and output but offer no mechanistic explanation[42]. This limitation means that in practical drug development, DNNs often serve only as screening tools, unable to provide actionable guidance for molecular optimization or mechanistic intervention.

To address this challenge, several efforts have emerged to enhance model interpretability[47]. The first category focuses on shallow interpretability enhancements, applying statistical attribution analyses over existing inputs. Common tools include SHAP and LIME, which quantify the contribution of input features (such as molecular weight, polarity, and topological indices) to the model's predictions. For example, some studies using SHAP analyses have identified high lipophilicity as a key factor influencing BBB permeability[48], while LIME analyses have shown how the presence of certain molecular fragments can significantly alter model outputs[49]. Additionally, attention mechanisms integrated into SMILES-based deep learning models enable the network to highlight important regions, such as specific aromatic rings, amine groups, and carboxyl groups. These methods are attractive for their ease of use and intuitive visual outputs, offering a degree of interpretability. However, they share a central disadvantage: they can only establish correlations between input features and predicted outcomes, without penetrating the deeper levels of transmembrane mechanisms, conformational dynamics, or local molecular interactions. As a result, the insights they offer have limited value in guiding practical molecular design.

The second category focuses on enriching the diversity and complexity of input information to increase model sensitivity to complex mechanisms[50]. Representative strategies include integrating multiple molecular representations (e.g., molecular graphs, SMILES encodings, and physicochemical descriptors), incorporating additional structural or pharmacokinetic

data—for example, three-dimensional structural information or membrane transport assay data—and employing multitask learning frameworks that jointly optimize BBB permeability alongside related pharmacokinetic or experimental measurements[51–53]. However, despite expanding the information space, these approaches largely remain forms of information integration and do not directly model molecular structure or mechanistic causality.

In contrast to these approaches, graph neural networks (GNNs) offer a fundamentally different modelling paradigm[54]. Rather than merely expanding inputs feature or stacking auxiliary labels, GNNs compute directly over molecular graphs, embedding the structural relationships into the computational process[28]. This advantage creates unprecedented opportunities to explore mechanistic variables, local fragments, topological features, and their relationships to functional outcomes[9], which offer a new lens in current interpretability frameworks[55]. To clarify how input representations influence model structure and interpretability, we summarize the comparative characteristics of various modelling strategies in Table 1. This transition—from low-dimensional descriptors to structurally and mechanistically enriched embeddings—marks a progressive redefinition of how permeability prediction is semantically formulated and structurally encoded.

**Table 1. Major molecular input representation paradigms in BBB permeability modelling and their structural and mechanistic characteristics.**

| Representation Category | Representative Encoding | Information Representation Level | Mechanistic Information | Key Limitations |
|---|---|---|---|---|
| Descriptor-based | Physicochemical descriptors; molecular fingerprints | Global physicochemical properties | Minimal | Low-dimensional; heuristic-driven; limited structural expressiveness |
| Sequence-based | SMILES strings; tokenized SMILES; SMILES embeddings | Linear symbolic sequence | Minimal | Order-sensitive; lacks structural constraints |
| Graph-based (2D) | 2D molecular graphs | Local molecular topology | Limited | No explicit 3D geometry; limited stereochemistry |
| 3D structure-based | 3D conformational graphs; atomic coordinates | 3D spatial molecular geometry | Structural | Conformer-dependent; computationally demanding |
| Mechanism-informed | Transporter/pathway tags; causal variables; mechanistic embeddings | Biological or mechanistic processes | Explicit | Annotation scarce; difficult to standardize |

The table compares common molecular input representation paradigms used in BBB permeability modelling. For each paradigm, representative encodings, the level of structural information captured, the degree of mechanistic information provided, and major methodological limitations are summarized.

### *II. Structural explanations via GNNs*

In response to the growing demand for interpretability, GNNs have emerged in BBB permeability prediction tasks as tools that[56] introduce explicit relationships between nodes, edges, and neighbourhoods, albeit unable to fully open the black box of prediction models[57]. This characteristic enables these models to at least partially analyze which molecular fragments, local structures, or subgraphs significantly influence the prediction outcomes[58]. Compared to traditional DNNs based on vector inputs, the special attribution capability of GNNs offers a new interpretive perspective[59], potentially layering the foundation for the future mechanism modelling[60].

The key breakthrough of GNNs lies in the design of their input layer: unlike DNNs, GNNs operate directly on molecular graphs, treating atoms and chemical bonds as nodes and edges[61]. Through methods such as graph convolutional networks (GCN) and graph isomorphism networks (GIN), they iteratively aggregate local neighbourhood information[62], integrating hierarchical signals from first-order neighbours (directly bonded atoms) to higher-order neighbours (fragments connected over multiple bonds). For example, under traditional input representations, molecules with identical formulas may still be difficult to distinguish if they differ only in the arrangement of functional groups[18], as fingerprint features often fail to distinguish such nuances. In contrast, GNNs can explicitly identify the spatial positions, chemical environments, and neighbourhood relationships of fragments like aromatic rings, amine groups and carboxyl groups[63], which effectively enhances the resolution of complex molecules, allowing the model to focus on specific local structural signals that are closely related to transmembrane transport or transporter binding. Moreover, beyond recognizing simple local groups, GNNs leverage multi-hop information propagation to comprehensively analyze complex arrangements and interactive effects[64] within molecules. For instance, in molecules with multiple adjacent hydrogen bond donors, GNNs can integrate the collective influence of these donors to identify potential transmembrane interaction networks, rather than merely calculating the contribution of individual donors. This refined capability allows GNNs to go beyond simply “identifying what the molecule is” to approaching “how the molecule internally operates,” offering entirely new perceptual channels for interpretability and subsequent optimization.

In practical applications, the capabilities of GNNs have been validated across multiple studies. The GraphADMET model, using multilayer graph convolutions and multi-hop neighbourhood aggregation, has been shown to detect the significant effects of minor functional group modifications on permeability[65]. Researchers found that simply introducing a methyl or hydroxyl group onto the main chain could lead the model to predict substantial changes in permeability. Notably, such subtle differences are often imperceptible to DNN models during the flattening into global features. MolGNet further incorporates graph attention mechanisms[9], which enables the model to automatically focus on key substructures such as cationic centers, aromatic backbones and flexible single bonds; even when multiple functional regions coexist within a molecule, the model can identify the areas that are the most sensitive to the task output. In a study[66], GNNs were combined with transporter activity predictions, and the results suggested that the model could not only recognize which fragments affect the overall permeability but also pinpoint potential binding sites[67]. For example, in certain molecules, the GNN successfully highlighted amino regions or ring structures likely to interact with *P*-glycoprotein (*P*-gp) transporters, providing directly usable, task-relevant clues for molecular optimization. This level of task-level interpretability and attribution not only enhances the comprehensibility of prediction outcomes but also endows GNN models practical utility in drug optimization[66], particularly in scenarios involving structural adjustments, site selection, and mechanism hypothesis validation.

Despite their notable advances in structural awareness and fragment attribution, the fundamental limitations of GNNs must be clearly acknowledged. First, the optimization objective of GNNs remains constrained to label fitting — whether in classification or regression tasks, the model ultimately learns statistical patterns between input features and output labels, not explicit mechanistic variables, causal pathways, or cross-scale processes. The model still cannot explain the detailed biophysical or chemical mechanisms through which local fragments influence permeability[19], nor can it simulate mechanistic changes arising from specific interventions (such as fragment removal, conformational locking, and surface modifications). Second, from a system performance perspective, GNNs place significantly higher demands on data scale, computational resources, and graph preprocessing compared to traditional DNNs. In small-sample or sparsely labelled settings, GNNs are prone to signal dilution and overfitting, while in large-scale high-throughput screening, their complex graph operations and multilayer aggregation mechanisms introduce considerable inference latency and resource consumption. Additionally, the robustness and generalizability of existing GNN architectures across different tasks and/or molecular spaces remain underexplored[68], limiting their widespread deployment in practical drug development workflows.

While GNNs have expanded the structured expression and local interpretability of molecular tasks[19], providing critical preconditions for mechanism modelling, further breakthrough will require more than structural attribution alone. The next stage of research must move toward mechanism-centered frameworks that bridge multiple scales and integrate multi-module collaborations, truly pushing molecular prediction tasks into a new phase of mechanism-driven, intervention-controllable and optimization-oriented modelling[69].

## From prediction to generation: An early leap toward mechanism-aware and integrative modeling

### *I. Redefining the task: from outcome prediction to mechanistic explanation*

In BBB permeability prediction tasks, discriminative modelling frameworks—particularly DNNs—have achieved significant progress, delivering robust performance across key metrics such as accuracy and AUC[70] and becoming important tools for high-throughput screening[71]. Despite these impressive results, such models often serve only as auxiliary references in real-world drug development rather than playing a central decision-making role[72]. The core limitation lies not merely in model performance but in the task definition itself: discriminative tasks focus on optimizing direct input-to-label mappings, without modelling the multilayer variables underlying molecular structure[73], transmembrane mechanisms, or transport pathways. As a result, the outputs provide statistical correlations rather than mechanistic explanations, failing to answer essential questions like “Which molecular features drive permeability?” or “How can structural modifications improve outcomes?”

Notably, the demand for mechanistic insight and structural guidance in drug development has existed for a long time[74] but could not be incorporated into task definitions due to multiple constraints: the lack of experimentally annotated mechanistic data, computational limitations[75] in handling complex structural and dynamic information, and industrial priorities that favoured efficiency over interpretability. In recent years, however, these barriers have gradually begun to ease, with the accumulation of experimental data on transporter activities and channel contributions, the enrichment of molecular graph and 3D conformation databases[76], and the rapid development of computational architectures such as GNNs, generative models, and causal inference frameworks[77]. Although current data volume and integration quality are still insufficient for fully mature mechanistic modelling, the conditions for exploration are increasingly coming into place.

The new task definition is no longer simply about predicting whether a molecule can cross the BBB, but about establishing causal chains between molecular structure[78], mechanistic variables, and functional outcomes. For example: Does the molecule cross via passive diffusion or active transport[79]? What functional groups, local conformations, or spatial arrangements play dominant roles[80] in the transmembrane behaviour? If a particular structural region is modified, how will it alter the mechanistic pathway and the ultimate permeability[3]? Such tasks require models to answer not only "what" but also "why" and "how to optimize[81]," providing actionable insights for drug design, screening, and refinement.

Therefore, "mechanistic modelling" is not merely about introducing more sophisticated algorithms—it represents a complete reshaping of the modelling framework[82]. It demands richer input layers (e.g., molecular graphs, 3D dynamic features, etc.), more complex output indicators (e.g., identification of mechanistic variables, inference of causal pathways, prediction of intervention outcomes, etc.), and more advanced optimization goals that integrate across modules[83]. This redefinition of task objectives directly drives the development of subsequent technologies—how to use structural awareness, latent variable learning, generative optimization, and causal reasoning to gradually break through the boundaries of traditional discriminative frameworks, opening new mechanism-driven paths for BBB permeability research. This redefinition of modelling objectives also reshapes how we categorize modelling approaches: no longer solely by algorithmic type, but by their capacity to encode, reason over, and act upon mechanistic variables. Figure 1 offers a comparative taxonomy across this evolving spectrum.

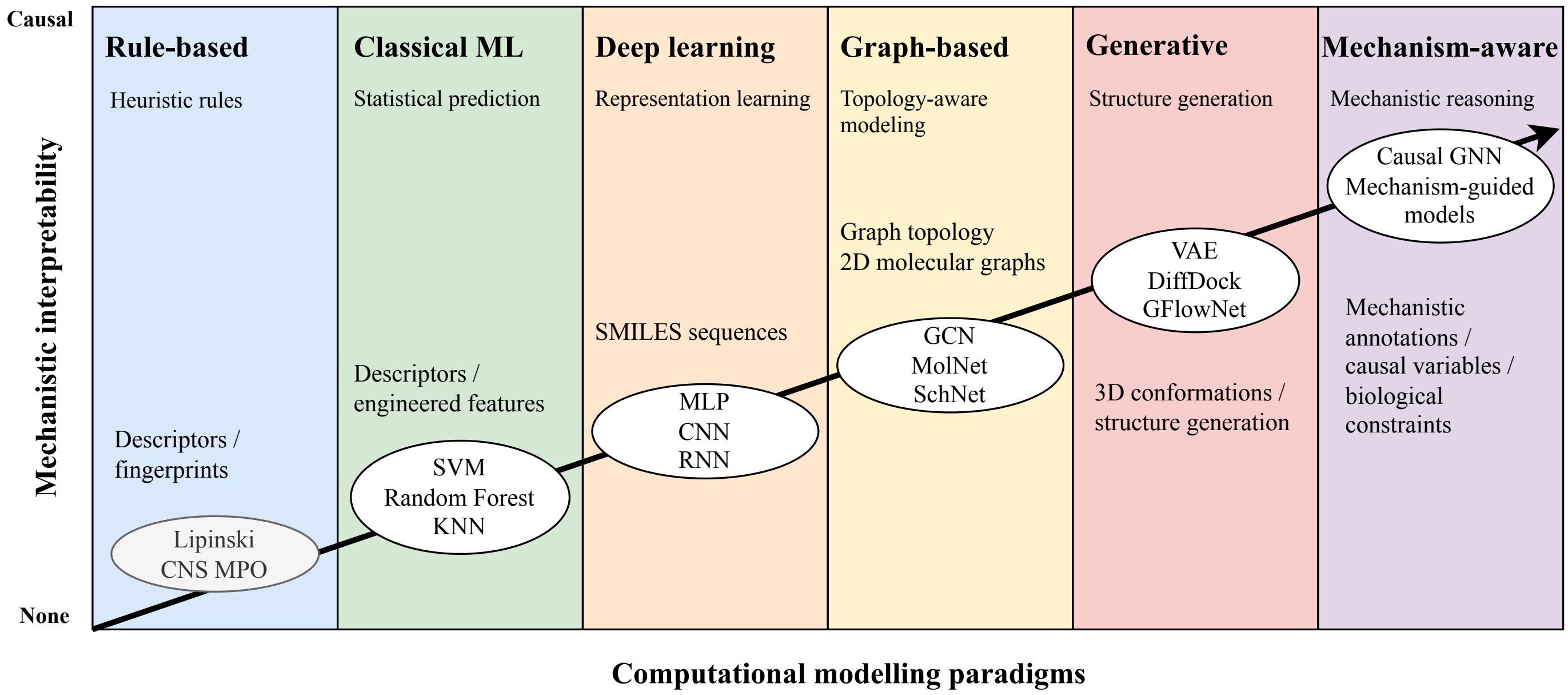

**Figure 1. Computational modelling paradigms for BBB permeability prediction across increasing levels of mechanistic interpretability.**
Modelling approaches are arranged by their dominant computational paradigm (horizontal axis) and increasing capacity for mechanistic interpretability (vertical axis). Representative models illustrate the conceptual transition from descriptor-driven statistical prediction toward structure-aware and mechanism-guided modelling.

## *II. Current mechanistic modeling approaches: foundational models and representational capabilities*

Current mechanistic modelling approaches for BBB permeability prediction cannot be understood as a single unified technique or algorithm[84]. Instead, they represent a multilayered system of interacting components, each designed to tackle specific challenges that arise when moving from purely statistical prediction to mechanism-aware reasoning. This layered architecture does not form arbitrarily but is driven by the inherent complexity of mechanistic modelling itself. To fully grasp it, we must unpack why mechanistic modelling demands distinct functional modules, how these modules are logically connected, and why they must work together as an integrated system.

At its core, mechanistic modelling does not merely aim to establish direct input-output mappings (e.g., predicting whether a molecule crosses the BBB) but requires the explicit introduction and handling of intermediate mechanistic variables—local structural features, dynamic behaviours, transport pathways[85], and/or even latent causal factors—that mediate the relationship between molecular structure and functional outcome. Achieving this goal cannot rely on a single modelling innovation; rather, it necessitates a progressive system where each layer builds up the system's capacity to represent, manipulate and exploit mechanistic information[86,87].

At the foundation, structural modelling components such as GNNs and their three-dimensional extensions (3D GNNs) provide the system with explicit molecular scaffolding. By representing molecules as graphs and applying message-passing mechanisms[88], GNNs capture local substructures and functional group combinations. 3D GNNs further incorporate spatial coordinates and conformational data[89], enabling the model to detect geometric and dynamic features beyond two-dimensional topology[90]. Without this structural foundation, the system lacks the capacity to localize or interpret mechanistic signals, making subsequent mechanistic modelling impossible.

Building upon this point, representation learning techniques such as self-supervised and graph contrastive learning[91] become essential for extracting latent mechanistic patterns, especially under conditions of limited data or sparse labelling[92]. These methods employ strategies like masking and reconstruction, positive-negative pair discrimination and cross-modal alignment, allowing the model to autonomously discover mechanism-relevant signals and construct richer[93], mechanism-sensitive representations. The enriched feature space, in turn, equips downstream components with the raw material necessary for mechanism-guided optimization and design.

Advancing further, generative modelling systems such as variational autoencoders (VAEs) and diffusion models[94] transform the system from a passive predictor into an active designer. Rather than merely evaluating known molecules, these models explore latent spaces, generating novel compounds that satisfy mechanistic constraints[95]. This behaviour represents a conceptual shift, reframing the task from screening to design, and enabling the model to contribute directly to hypothesis generation and innovation[96].

At the highest abstraction, causal inference architectures such as causal GNNs and structural causal models[96] embed explicit causal variables, pathways, and intervention reasoning. These models aim not only to learn correlations but to reconstruct causal chains, predict the outcomes of hypothetical interventions, and support mechanism-driven scientific inquiry. Although still at an exploratory stage, these methods represent a profound transition from statistical learning to causal reasoning[97].

Importantly, the layers are not independent or interchangeable; they form a progressive chain instead. Structural encoding enables mechanistic representation[98]; representation learning enriches generative exploration[99]; generative models supply objects and hypotheses for causal reasoning. The true future breakthroughs in mechanistic modelling will rely not only on advancing individual modules but on integrating them into closed-loop systems, connecting molecular structure, mechanistic understanding, design, and intervention[100]. Such integrated frameworks hold the potential to transform not only BBB permeability prediction but also the broader field of mechanism-driven drug discovery[101], opening new frontiers in both scientific insight and practical application.

## *III. Toward Practical Implementation: From Mechanistic Modelling to Real-World Application*

While the redefinition of the task provides a solid conceptual foundation, realizing its potential in practice demands operational clarity across four key dimensions: data, modeling strategy, evaluation, and application. Mechanistic modeling requires data that capture the causal and structural determinants of BBB transport—such as transporter interactions and conformational flexibility—but such annotations are still limited. Hybrid strategies like simulation-based augmentation and curated transporter knowledge can help embed these mechanistic signals while reducing superficial correlations.

Mechanism-aware modelling should reflect the hierarchical and causal organization of biological systems[102]. Instead of relying on black-box mappings from molecular descriptors, models should adopt a compositional architecture aligned with mechanistic structure[103]. At the base, multi-task learning leverages auxiliary signals—such as transporter affinity and subcellular localization—to guide the extraction of disentangled, mechanism-relevant features[104]. Causal modelling builds upon this backbone by introducing intermediate variables (e.g., rate-limiting steps, transcytotic routes, etc.), enabling structured reasoning beyond input–output fitting[105]. To address data heterogeneity, multi-fidelity models fuse signals across resolutions—from low-precision estimates to experimental or simulation-derived measurements—ensuring robustness under uncertainty[106]. At the highest layer, physics-informed neural networks (PINNs) incorporate known biophysical laws (e.g., Fick's diffusion, electrochemical constraints, etc.) directly into training, reinforcing mechanistic grounding[107]. Together, these layers form an integrated scaffold—from supervision to structure and then to constraint—that elevates prediction into biologically coherent reasoning. These mechanistic constructs provide a testable substrate for evaluating causal fidelity in BBB transport modeling.

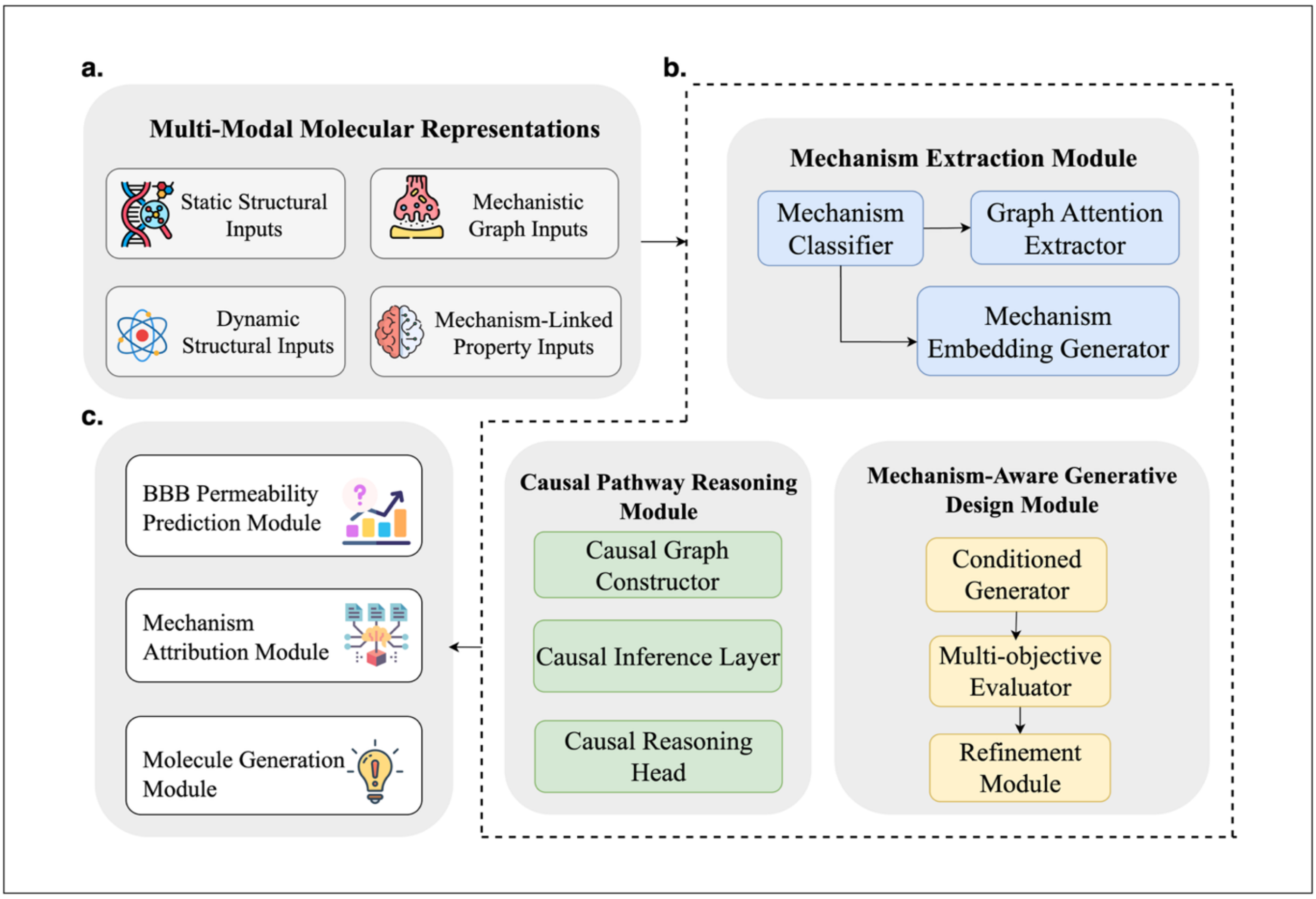

**Figure 2. A unified framework for mechanism-aware modelling of BBB permeability and molecular design.**

(a) Multi-modal molecular representations integrate static/dynamic structural features, mechanistic graph inputs, and property-linked descriptors to enrich input diversity.

(b) The mechanism extraction module encodes mechanistic information from inputs via a classifier, graph attention extractor, and embedding generator.

(c) Three task modules operationalize mechanism-aware modelling: (1) BBB permeability prediction, (2) mechanistic attribution, and (3) mechanism-guided molecular generation.

To truly advance BBB modelling, evaluation must shift from quantifying prediction accuracy to determining whether a model has internalized the causal logic of barrier transport[7]. Performance scores such as ROC-AUC or RMSE can certify statistical fit, yet remain silent on whether predictions are driven by physiologically meaningful signals or by spurious dataset correlations[15]. A mechanism-aware evaluation therefore treats the model as a putative biological system: do its predictions reconfigure when key physicochemical determinants are perturbed? Do its internal explanations converge on established transport rules rather than incidental descriptors? And crucially, when mechanistic pathways are disabled—such as simulating transporter knockouts or competitive inhibition—does the model's behavior reorganize in the expected direction, revealing dependence on actual biological routes rather than mere pattern recognition? Validation reaches its highest standard when models not only reflect mechanisms but deploy them as functional tools—actively proposing structural modifications that enhance receptor-mediated uptake, reduce efflux liability[108], or diversify transcytotic solutions while maintaining drug feasibility. In this progressive view, evaluation becomes a stress test of mechanistic alignment: from perturbational responsiveness to mechanistic attribution, to causal compliance, and finally to operational design capability. Only by passing through this hierarchy of evidence can a BBB model be said to embody mechanism as computation—transforming predictive insight into actionable guidance for CNS drug discovery.

Mechanism-aware BBB models are valuable not because they predict permeability, but because they inform how to improve it[109]. By representing explicit mechanistic variables—diffusion, transporter engagement, efflux susceptibility—the model can trace which structural features drive barrier behavior, revealing where targeted modifications would most effectively shift transport outcomes. Counterfactual simulation then extends this reasoning into actionable design: adjusting polarity, modulating rigidity, masking hydrogen-bond donors, or simulating P-gp inhibition allows in silico exploration of transport route manipulation before synthesis. When deployed at scale, the same mechanistic constraints enable generation and prioritization of candidates with BBB-compatible profiles that remain medicinally feasible—supporting scaffold innovation while reducing experimental burden[110]. Thus, what begins as mechanistic prediction becomes strategic guidance, turning BBB modelling into an engine for rational CNS lead optimization rather than a retrospective classifier.

**Current limitations and future directions: challenges and paths toward integrated mechanistic systems**

DNNs remain the dominant approach for BBB permeability prediction, balancing strong performance with ease of deployment[111]. Yet their black-box nature limits insight into the structural or causal factors governing trans-barrier behavior. GNNs offer improved molecular representations via structural encoding[18], but remain constrained by discriminative, correlation-based paradigms—falling short of capturing the biological mechanisms underlying permeability.

These limitations have made increasingly clear that performance alone is not enough. As predictive models move closer to practical deployment, the emphasis must shift from surface-level correlations to mechanistic interpretability. This need has motivated a growing turn toward mechanism-aware modelling[112]—a paradigm aiming not merely at forecasting outcomes but at reconstructing the causal chains linking molecular structure, transport kinetics, and physiological function. Unlike DNNs and GNNs, which primarily learn statistical associations, mechanism-aware approaches explicitly model intermediate variables such as transporter affinity, conformational transitions, and pathway-specific dynamics. These embedded mechanistic representations enable models to explain why a compound is permeable[41], predict how structural changes may modulate its behavior, and generate actionable insights to guide rational molecular design. Grounding prediction in mechanistic understanding will also enhance generalizability across diverse chemical scaffolds, ensure biologically coherent interpretability, and support feedback-driven optimization cycles—capabilities essential for achieving translational impact.

Current mechanistic frameworks remain in the early stages of development and are faced with several persistent challenges, including the scarcity of well-annotated mechanistic datasets, fragmented modelling pipelines, and the absence of unified integration strategies[113]. While foundational tools—such as generative models and causal inference architectures—are beginning to emerge[41], they have yet to form a cohesive system capable of guiding real-world molecular design. Moving forward, the goal is not to replace existing DNN-based infrastructures, but to augment them through a multi-layered, closed-loop architecture that integrates structural encoding, mechanism-specific representation learning, generative molecular design, and causal inference[68]. Such systems will not only enhance predictive accuracy, but also support hypothesis generation, design optimization, and mechanistic interpretability—transforming BBB modelling from statistical forecasting into intervention-guided design[114]. More broadly, this transition reflects a deeper shift in drug discovery itself: from learning patterns to modelling principles, and from data-centric predictions to biologically grounded reasoning. This transition, as outlined throughout this review, reflects not only a technical trajectory but a redefinition of how predictive models are framed, evaluated and deployed in neuropharmacological discovery.

**Acknowledgments**

The authors gratefully acknowledge funding from the Westlake Fellows Program at Westlake University.

**Competing Interests**

The authors declare no competing interests.

**Data and Code Availability**

No original data or code are associated with this review.